\documentclass[trackchanges,twocolumn]{aastex701}

\usepackage[english]{babel}

\usepackage{amsmath,amssymb}
\usepackage{graphicx}
\usepackage{threeparttable}
\usepackage{multirow}
\usepackage{threeparttable}
\usepackage{bm}
\usepackage{tabularx}
\usepackage{tikz}
\usepackage[dvipsnames]{xcolor}

\begin{document}

\title{Multimessenger Constraints on the Central Engine in SN 2023ixf: Neutron Star or Black Hole?}


\author[0000-0002-1518-1946]{Maryam Aghaei Abchouyeh}
\affiliation{Department of Physics and Astronomy, Sejong University, 98 Gunja-Dong, Gwangjin-gu, Seoul 143-747, Republic of Korea}
\email[show]{} 

\author[0000-0002-9212-411X]{Maurice H.P.M. van Putten}
\altaffiliation{Corresponding author, mvp@sejong.ac.kr}
\affiliation{Department of Physics and Astronomy, Sejong University, 98 Gunja-Dong, Gwangjin-gu, Seoul 143-747, Republic of Korea}
\affiliation{INAF-Osservatorio Astronomico di Capodimonte, Salita Moiariello 16, I-80131 Napoli, Italy}
\email[show]{} 

\author[0000-0003-3142-5020]{Massimo Della Valle}
\affiliation{INAF-Osservatorio Astronomico di Padova, Vicolo dell’Osservatorio 5, 35122, Padova, Italy}
\email[show]{} 

\begin{abstract}
{The physical trigger of core-collapse supernova (CC-SNe) is a central theme 
in modern surveys of the transient Universe.
A multi-messenger approach may provide
a unique window to CC-SNe 
produced by a newly born neutron star (NS) or black hole (BH).}
Here we {apply joint electromagnetic-gravitational wave (EM-GW) analysis} to the nearby event SN 2023ixf 
{using a model-independent broadband search with a 
GW-energy detection threshold of a few times $10^{-5} M_\odot c^2$ for chirps and $10^{-2}$ $M_\odot c^2$ for noise-like emission below 1\,kHz.}
{No joint LIGO Hanford-Livingston (H1-L1)} signal associated with SN 2023ixf is observed
over a 2.5 day window proceeding the first discovery. 
Together with {EM} photometric-spectroscopic data, 
we interpret this null result to indicate the absence of BH formation, in part by mass-scaling of the BH powering GW170817B following GW170817. 
This {multi-messenger} outcome supports, hereby, 
scenarios consistent with NS formation, 
{consistent with a progenitor mass likely below $\sim 20 M_\odot$ estimated by EM interpretations.}

\end{abstract}

\section{Introduction}
\label{S_Intro}

{Core-collapse supernovae} (SNe) are one of the most extreme transients in the universe powered by a newly born NS or BH \citep{Lipunov2007,Tonry2018,Graham2019,Jones2021,Aleo2023}. 
While observed at increasingly large numbers, the nature of their remnants, mass of their progenitor stars, and the physical mechanism powering these events, are still mysterious \citep{Janka2012,Sawai2014,Papish2015,Pejcha2015,Oberg2020}. This is dramatically demonstrated by SN1987A whose light curve in {electromagnetic (EM)} and $\nu$ emission were notoriously insufficient to determine the remnant \citep{Alp2018}. Its remnant has only recently been found to be most likely a NS \citep{Fran2024}
\citep{Gilm1987,West1987,Kish1987}.

Identification of central engines and remnants of CC-SNe are tightly correlated with the progenitor mass. Progenitors of mass $8M_\odot \lesssim  M\lesssim 20M_\odot$ are {generally} understood to produce core-collapse supernovae, (CC-SNe) associated with the formation of NS \citep{Smartt2009,Sukhb2016} 
and those of mass $20M_\odot \lesssim M$ are expected to produce a (rotating) BH, even as they may briefly pass through the phase of a hyper-massive NS (HMNS) for a finite time (e.g. \cite{Izzo2012}). 
Both groups are potentially luminous in all three channels of EM, $\nu$ and {gravitational waves (GW)}, {where the MeV$\nu$ emission is relatively secure for the NS case according to SN 1987A. Any gravitational radiation is uncertain due to detailed dependency in multiple mass moments \citep{Smartt2009}.} 
We note that observational evidence and theoretical arguments are compatible with this scenario, indicating that many stars above about $M\gtrsim20M_\odot$ may undergo direct collapse to black holes, in some cases without producing an observable {supernova (SN)} (e.g. \cite{Fryer1999,Belc2012}).

The physical trigger of CC-SNe plays a central role in the energy emitted in different radiation channels during the event, hence identification of progenitor and central engine. Observational shortcomings of neutrino driven CC-SNe \citep{Janka2012}, shifts our focus on physical triggers to accretion/angular momentum in powering CC-SNe by magneto-rotational mechanisms \citep{Bis1970,Bis1971}. {Aligned with the angular momentum of a central energy reservoir}, these mechanisms guarantee a preferred direction consistent with the observed SNe and their extended supernova remnants (e.g. \cite{Yang2025}). 

{$E_J$, the rotational energy, is central to this scenario for both NS and BHs. For NS, $E_J=\tfrac12I\Omega^2$ follows the conventional expression given its moment of inertia $I$ and angular velocity $\Omega$ below the break-up angular velocity. For a BH of mass $M$, $E_J$ follows the Kerr solution satisfying $E_J\lesssim 0.29Mc^2$. A key difference between the two is that the $E_J$ of a BH can exceed of that of a NS by 1-2 orders of magnitude \citep{mvp2023}.}
$E_J$ can be radiated in different channels mentioned above 
covering EM, MeV neutrino, GW, and possibly an ultra-relativistic baryon poor jet, where GW radiation tends to be the dominant radiation channel in BH powered SNe.  

Notably, black holes exhibit remarkable universality {by} scale-invariant relations across different mass scales in astrophysical outflows from active galactic nuclei (AGN) to microquasars \citep{mer03,fen03,fal04,fen12}. 
By the no-hair theorem, this universality also implies that black holes, once formed, have no memory of their progenitor {system}, apart from total mass and angular momentum. {Consequently}, 
any ensuing radiation is hereby set by initial mass and spin of the newly formed BH, regardless of their progenitors, i.e., {compact binary mergers including NS-NS or NS-BH mergers, or core-collapse of massive stars} \citep{mvp2024}. 
{The common gravitational radiation signature hereof is a long duration spin-down chirp. \textcolor{black}{This GW emission is apparent in a descending signal GW170817B,} identifying the central engine of GW170817B-GRB 170817A to be a rapidly spinning BH \citep{ligo2017,mvp2023,ligo2021}. Any accompanying EM radiation may be significant to event timing, but not pertinent to the ultimate prospects of a GW detection.}

{This universal behavior of} black holes powering CC-SNe implies an observational horizon distance $D\simeq 160\,$Mpc to gravitational radiation during {LIGO's fourth observational run O4} \citep{mvp2024,Cutler2002}, significantly beyond a few Mpc, typically quoted for 
LIGO searches for CC-SNe 
{based largely on NS central engines or quasi-normal mode ringing (QNMR) of newly formed BHs, both at high frequencies significantly above LIGO bandwidth of sensitivity.} \citep{Abbott2020,szc2021,Abac2025}.

{
Accordingly, the core-collapse supernova SN 2023ixf at $D\sim 7$\,Mpc provides a fortuitous opportunity to probe its physical trigger by gravitational-wave observations,
taking advantage of advances across several fronts that have fundamentally changed what is achievable with present
technologies.} {This includes improvements in LIGO sensitivity by about an order of magnitude over the last decade that can now be fully exploited for broadband searches employing template banks of size $\mathcal{O}(10^6)$ using high-performance computing. 
It thus becomes feasible to place meaningful astrophysical constraints on central engines of CC-SNe, beyond exploratory analysis of the type Ib/c event SN~2010br 
($D\simeq8\,{\rm Mpc}$ in NGC 4051) during the {sixth science run of LIGO-I (S6)} \citep{van16A}. 
For the recent SN~2023ixf, therefore, the combination of the two independent observational channels of EM and GW promises to provide powerful direct constraints on whether its central engine settled into a neutron star or collapsed to a black hole.}

{To this end, we perform a model-independent deep {GW} search at detector-limited sensitivity for any potential joint signal in {the Hanford-Livingston detectors H1 and L1}, extended over all available data covering 2.5 days before the first detection with no additional assumptions in our data analysis. The results are then combined with spectroscopic-photometric EM outputs.}


In \S \ref{S_ixf} we discuss priors on {potential NS or BH central engine of SN 2023ixf}. In \S \ref{S_prerequisite} we review our image-based model-agnostic and broadband GW-search pipeline, followed by our observational results. Following an in detail discussion in \S \ref{S_discussion}, we conclude our results in \S \ref{S_conclusion} indicating the central engine of SN 2023ixf within the availability of GW data.

\section{On the central engine of SN2023\lowercase{ixf}}
\label{S_ixf}

SN 2023ixf is first reported on May 19th, 2023 at 17:27:15.000 \citep{Itagaki2023}. Further archive searches show earlier detections by other groups with the earliest 
on the same day at 03:41:35.000 \citep{Limeburner2023}. SN 2023ixf is classified as a {SN type II} and its progenitor star is estimated to be a red supergiant \citep{perley2023,Kilp2023,Chandra2024}. This event is observed at a distance of $D\simeq 6.85\pm0.15$\,Mpc \citep{Riess2022,Jenson2023}, {well within the 160\,Mpc horizon distance} (\S \ref{S_Intro}). By its proximity, SN 2023ixf is the closest SN observed in the last decade, 
which gives a good opportunity to search for any potential GW signal associated with this event. 

Based on the mass loss of the progenitor star few years prior to SN 2023ixf, studies of the properties of the surrounding environment, hydrodynamical models and periodic variability of SN 2023ixf, the mass of the progenitor has been estimated to be within $12M_\odot<M<22M_\odot$, though the low and high limits are not strongly supported by model-independent observations \citep{Sora2023,Bers2024,Ferr2024,Fola2025}. Notably, this mass scale coincides with the empirically inferred upper limit for red-supergiant progenitors of Type II supernovae, estimated to be $\sim 17–21 M_\odot$ from statistical studies of pre-explosion detections \citep{Smartt2009,Ben2018,Ben2020,Koch2020}. \textcolor{black}{Therefore, EM provides constraints for the progenitor mass of SN 2023ixf, but much less so for the central engine}. 

{GWs on the other hand, can be conclusive by breaking the degeneracy between a BH and NS central engine considering different $E_J$ each can accommodate. By the} universality of BHs and the no hair theorem, gravitational radiation from a BH central engine powering SN~2023ixf is expected to be a descending GW-chirp, similar to that of GW170817B, but with significant difference in observed strain, primarily due to its proximity and, to a lesser extent, an expected broad range of potential BH mass. 

GW170817B is a {spin-down} signal observed at $5.5\sigma$ confidence level 0.92 seconds after the ascending chirp signaling the merger of two NS at 40Mpc. The gravitational energy emitted during the descending signal is measured to be $E_{GW}\simeq 3.5\%M_\odot c^2$ over the frequency range $\sim 700$\, Hz to $\sim 200$\, Hz. This signal is further supported by the consistency of the independent total mass estimates for these GW170817 and GW170817B chirps. {$E_{GW}$ is approximately 2-3 times the maximum rotational energy of a NS at the corresponding mass and gravitational wave frequency. 
As the BH central engine released its rotational energy in {GW170817B-GRB 170817A-AT2017gfo}, hence spinning down during this complex radiation process in interaction with surrounding high-density matter \citep{mvp2023}.}

\begin{figure}
    \centering
    \includegraphics[width=1\linewidth]{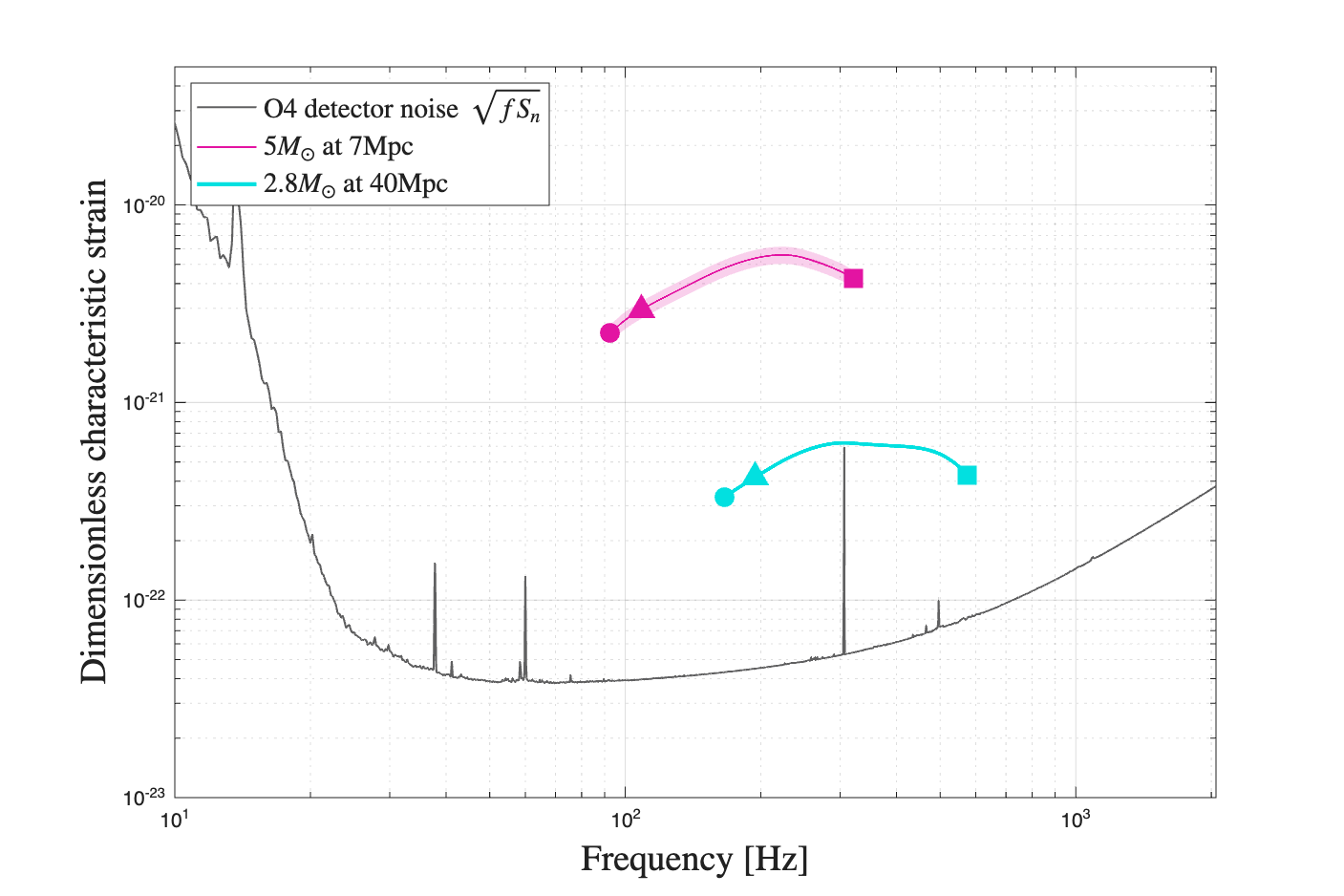}
    \caption{ 
    Detectability of descending GW chirps from BH spin-down during O4. The curves show the characteristic strain for different BH masses and source distances, including SN 2023ixf ($\sim 7$ Mpc, magenta) and GW170817B (40 Mpc, cyan). Square and triangular markers indicate initial BH spin values $\hat{a}=0.8$ and $\hat{a}=0.5$ respectively, while filled circles mark the theoretical termination of the signal at $\hat{a} \sim 0.44$. Quite generally, the detectability of a BH spin-down with initial spin $\hat{a}<0.8$ follows the same curve starting from commensurably lower frequencies. The curves show the the result for theoretical orientation-averaged signals based on ideal matched filtering \citep{Flan1998,Cutler2002}, which is slightly higher than what is realized in BMF.}
    \label{fig:spectrum}
\end{figure}

Regardless of any model, SN 2023ixf is a CC-SNe with a potential BH central engine, observed at a distance closer than GW170817B by a factor of about $k_d\simeq 6$, during {the 15th engineering run ER15} prior to O4 with sensitivity $k_D \simeq 1.8$ times that of O2. Additionally, the mass of BH remnants of CC-SNe are more likely to be $\gtrsim 5M_\odot$ making the event more energetic by a factor of $k_m\simeq 2$, moving the signal closer to the most sensitive frequency band of LIGO, adding another improvement factor of $k_n\simeq 1.3$ to the previous ones \citep{mvp2024}. Combined, these independent improvement factors show that a descending GW chirp from a BH central engine of a CC-SNe observed during O4 {at this distance} is observationally $\mathcal{K}$ times stronger than GW170817B. Measured by the anticipated SNR, we define $\mathcal{K}$ to be the enhancement in {signal-to-noise ratio (SNR)} over GW170817B, satisfying

\begin{equation}
    \label{EQN_k}
    \mathcal{K}=k_d\times k_D\times k_n \times k_m\simeq 28.
\end{equation}

\noindent Given the SNR $\sim 6$ of GW170817B \citep{mvp2019}, {equation \eqref{EQN_k}} increases the SNR to the incredibly high value of about 170. 

{Fig. \ref{fig:spectrum} shows the detectability by dimensionless characteristic strain during O4 of such signal alongside the modeled descending chirp of GW170817B, previously observed during O2. Inspired by GW170817B and for illustrative purposes, {we consider $\hat{a}\simeq 0.8$ as the initial {normalized Kerr parameter, $\hat{a}=c J/GM^2$ for a BH of mass $M$ and angular momentum $J$}. Due to the proximity of SN 2023ixf (cf. {equation \eqref{EQN_k}}) the spin-down} feature is projected to be detectable across a broad range of black hole masses. 
    
Therefore, any descending chirp associated with SN 2023ixf, if present, would be expected to produce a clearly identifiable signature in the data given the enhancement factor $\mathcal{K}$ in {equation \eqref{EQN_k}} using a model-independent data analysis method. {Not observing such signal in GW data analysis is} consistent with a scenario in which the central engine is an NS. }

Note that $E_J$ is about two orders of magnitude smaller for NS central engines, decreasing the SNR and hence detectability by the same order of magnitude. Therefore NS central engines tend to be non-observable or marginally observable at best.

    \begin{figure}
        \centering
        \includegraphics[scale=0.33]{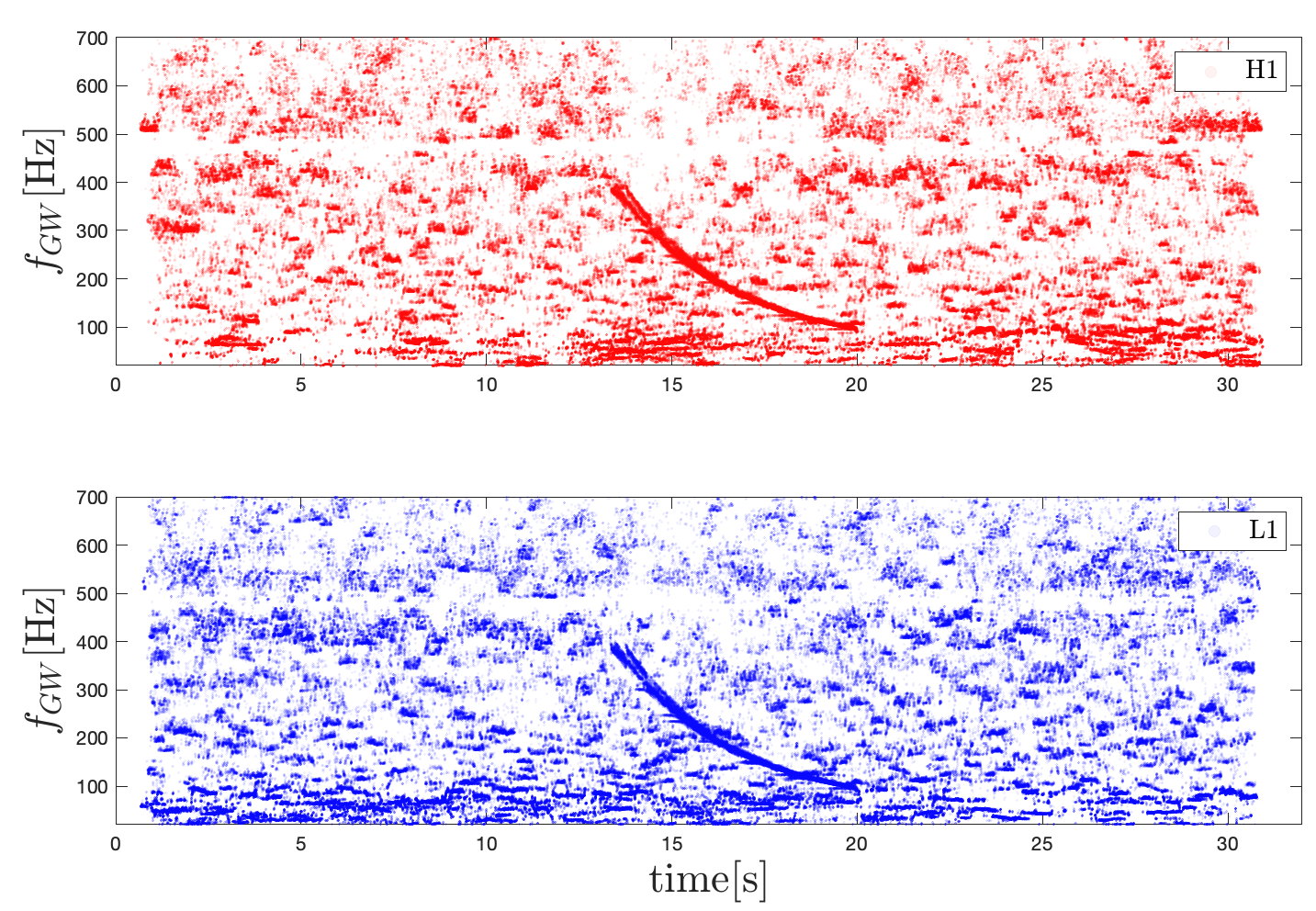}
        \caption{ Simulated long-duration descending GW chirp expected from a rapidly rotating black-hole central engine at a distance of 7\,Mpc, here injected to O4 LIGO data for an initially rapidly spinning BH of mass $M\simeq 5M_\odot$. The signal is created by mass-scaling of the black hole central engine of GW170817B with initial Kerr parameter $\hat{a}\simeq 0.8$.
        As shown, it results in a long duration descending GW chirp spanning frequencies from about 400Hz down to about 100Hz. 
        No comparable feature is observed prior to the optical onset of SN 2023ixf. Within this framework the null detection disfavors such engines and is instead consistent with a NS central engine and remnant.}
        \label{fig:BMF}
    \end{figure}


\section{Deep GW probe of SN 2023ixf}
\label{S_prerequisite}

In searches for GW emission from SN 2023ixf, we here follow the road-map and expectations outlined previously in \cite{mvp2024}.

 In contrast to merger signals, CC-SNe events do not have a clear event trigger time, $t_0$, unless associated with a {gamma-ray burst (GRB)} which reduces the uncertainty in $t_0$ to seconds. For a usual CC-SNe without a GRB, the first trigger time $t_O$ (observed time) is later than shock break-out time $t_b$ of the order of hours to weeks, and the core-collapse time $t_0$ is earlier than $t_b$ by hundreds of seconds to hours depending on the progenitor star. With these uncertainties, estimating $t_0$ would be based on statistics and EM observations. For SN 2023ixf, $t_0$ is estimated to be about a day before the first detection \citep{Hossein2023,Kozy2025}.

Having the core-collapse time at hand, we do a model-independent image based search on LIGO data around this event using Butterfly Matched Filtering (BMF), first introduced in analyzing the GRB light-curves to identify their {broadband turbulent spectrum at high frequencies, subsequently applied to GW strain data \citep{mvp2016}. Performed using modern {high-performance computing (HPC)}, BMF can be further accelerated by signal processing extended to wafer-scale engines with $\mathcal{O}(10^6)$ processing elements \citep{van25D}. 
}

\textcolor{black}{{By definition, BMF is} performed on a very dense bank of time-symmetric templates, covering a broad range of frequency and time rate-of-change of frequency. Detecting the Kolmogorov spectrum of turbulence \citep{mvp2014b} and both the ascending and descending branches of
GW170817 and, respectively, GW170817B, provides evidence for the sensitivity of BMF to a broad range of model-independent signals. Therefore BMF has {\it no bias} toward the shape or monotonicity of the signal.} 

On the other hand, the detection of {full} GW170817-GW170817B shows that BMF does GW search at near detector-limited sensitivity. Therefore, by construction, BMF gives the opportunity for a model-independent and deep search for all types of unknown signals associated with CC-SNe. {Note that BMF detects GW170817B 
independently of GW170817, i.e. 
these two detections are mutually blind to each other 
by data analysis and theory of radiation processes.}

In practice, working with O4 data, BMF provides a broadband search algorithm with a gravitational-wave energy thresholds of the following two types

\begin{eqnarray}
\begin{array}{lcccc}
   \mbox{Chirp (ascend/descend):} &\mathcal{E}_{th,GW}^{chirp} \simeq &\mathcal{O}(10^{-5}) M_\odot c^2\\ \\
   \mbox{Noise-like (broadband):} &\mathcal{E}_{th,GW}^{bn}\simeq &\mathcal{O}(10^{-2}) M_\odot c^2,
\end{array}
    \label{EQN_threshold}
\end{eqnarray}

\noindent for emission at frequencies $\lesssim 1$ kHz. These detection energy thresholds are equivalent to horizon distances of about 160 Mpc for chirps and 5 Mpc for noise-like signals, respectively.

{Here we apply BMF to the data covering the time window $[t_0, t_O]$ of SN 2023ixf, following the ground} rule that a candidate signal must be observed in both L1 and H1 detectors \citep{Abc2023}. This is especially important in search for new and unknown signals.

Generally, a search for a GW signal (jointly in H1 and L1) from a CC-SNe at low frequencies, i.e. $30 \lesssim f_{GW} \lesssim 1000$Hz, may come to three {alternative outputs \citep{mvp2024}: (i) positive detection, (ii) a confident null detection or (iii) a  signal from an imposter}, i.e., {another event that coincidentally appears associated with the event, but is, in fact, unrelated}. These respectively point to a BH central engine, a confident NS central engine, and a slowly rotating NS central engine or a BH spin-down of another event which accidentally lies within the search window of SN 2023ixf.

Given the estimated $t_0\simeq 1$\,day, we searched about 2.5 days of 16kHz sampling rate data before the first event trigger time (covering about 14 hours after the earliest detection - \S \ref{S_ixf}) within the availability of GW data using spectrograms of 32 seconds. These spectrograms were inspected over a frequency range of 30-1000 Hz for a joint observation in two detectors. BMF spectrograms are then calculated and plotted (see Fig. \ref{fig:BMF} for an example) for the entire data in use, giving $\mathcal{O}(10^4)$ spectrogams. 
Considering the origin of the present candidate signals being from high-density matter in a state of magneto-hydrodynamical turbulence, this intermediate time-scale of phase coherence is essential to capture GW chirps of otherwise general form. In our image-based search, the resulting spectrograms, partitioned over segments of 32 s, are screened for any joint signals of duration $T>\tau$ appearing simultaneously in both H1 and L1. 


\begin{figure}
    \centerline{\includegraphics[width=1.2\linewidth]{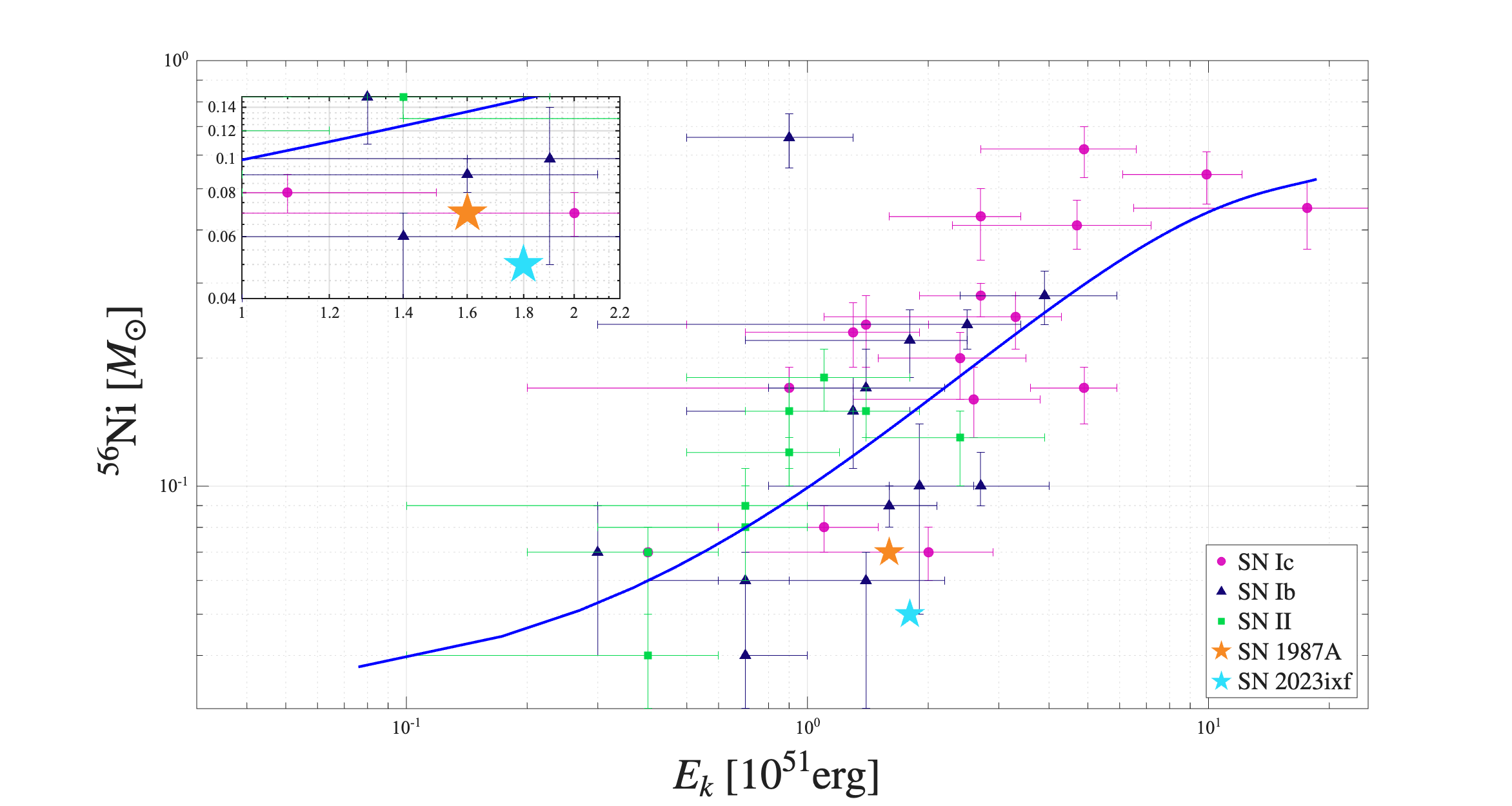}}
    \caption{
    {Correlation of $^{56}$Ni with kinetic energy $E_k$ for 38 stripped SNe discussed in \cite{Lyman2016} and \cite{mvp2024}. 
    The sample includes Type Ib, Ic and II SNe showing more energetic SNe to produce greater amount of $^{56}$Ni. To guide the eye, shown is a cubic fit $y=7\times 10^{-5}x^3-0.0037x^2+0.07x+0.032$ (blue curve).
    {(Insert.) SN~1987A and SN~2023ixf} (orange and cyan stars, respectively)
    are both relatively less energetic events with similar $E_k$ and $^{56}$Ni.}
    }
    \label{fig:Mass}
\end{figure}

{To illustrate this according to {equation \eqref{EQN_k}}, we performed an injection experiment for a potential BH spin-down associated with SN 2023ixf. Fig. \ref{fig:BMF} shows the result for a fiducial BH mass $M=5M_\odot$. The resulting signature in the spectrogram is indeed {\it impossible to miss}.}

{Notably, we do not assume any assumption on the potential central engine.}

\section{Results}
\label{S_discussion}

Screening the available data over a window of about 2.5 days before the first detection of SN 2023ixf, our general search produces a null result with no detection of ascending, descending nor broadband noise signal above the thresholds in {equation \eqref{EQN_threshold}}.

A more complete picture on the central engine and progenitor appears considering also the kinetic energy $E_k\simeq (1.8\pm0.2) \times 10^{51}$\,erg in mass-ejecta $M_{ej}\simeq 9M_\odot$ at velocity $v_{ej}\simeq 5 \times 10^{8}{\rm cm\,s^{-1}}$ {with $M_{^{56}{\rm Ni}}\simeq 0.05\,M_\odot$ \citep{Teja2023,Mori2024,Kozy2025}} associated with SN 2023ixf. {We here interpret our results based on a multi-messenger approach} including gravitational-wave observations and these photometric-spectroscopic constraints on the SN 2023ixf, typical for type II SNe \citep{cos25}.

\subsection{Black hole central engine?}
Quite generally, radiation from BH central engines can be spin powered and/or accretion powered. 

Considering the {null result following a search in GW}, the low detection threshold for descending branches in {equation \eqref{EQN_threshold}} notably rules out {black hole spin-down, practically} at any choice of black hole mass or initial spin (Figs. \ref{fig:spectrum}-\ref{fig:BMF}).

For the possibility of {(hyper-)accretion onto} a black hole -- generally causing BH spin-up \citep{Bardeen1970} -- we note the non-detection of broadband noise-like gravitational-wave {emission about threshold equation \eqref{EQN_threshold}}. 
However, the gravitational energy radiated during (hyper-)accretion is estimated to be about $E_{gw} \simeq 10^{-5}-10^{-4}M_\odot c^2$ \citep{Gott2024}. {This is far below the $10^{-2}M_\odot c^2$ detection energy threshold in equation \eqref{EQN_threshold}. Therefore, this 
gravitational-wave signature of accretion flows appears insufficient to be confidently constrained by 
gravitational-wave observations alone. }

If the above-mentioned $E_k=\left(1.8\pm0.2\right)\times 10^{51}$\,erg nevertheless derives from accretion flow shedding angular momentum in a magnetic wind with energy $E_w$, this process tends to be inefficient given the non-relativistic ejection velocities $\beta_{ej}=v_{ej}/c\simeq 0.017$. Powering $E_k=\eta E_w$ at $\beta_w=v_w/c$ is governed by the efficiency $\eta \sim (1/2)\beta_{ej}$ \citep{van11}, giving $E_w\lesssim 2.4\times 10^{53}$\,erg. This bound refers to baryon-poor relativistic winds ($\beta_w\simeq 1$) -- perhaps consistent with GRB-SN scenarios {but not SN 2023ixf at hand}. 
Deriving $E_k$ from more efficient, baryon-rich winds with $\eta \sim \beta_{ej}/\beta_{w}$ where $\beta_w \ll 1$, would require a substantial mass loading of a few $M_\odot$, possibly over-producing the modest amount $^{56}$Ni typical for type II events and well below what is produced in their more energetic counterparts \citep{Smartt2009}. 

{By these multi-messenger observational constraints, a black hole central engine powering $E_k$ by its spin-energy or by (hyper-)accretion
appears unlikely.}

\subsection{Neutron star central engine?}
{The above mentioned observational constraints on $E_k$ and $^{56}$Ni for SN2023ixf} are consistent with that of typical type II supernovae and in particular also with SN 1987A \citep{Seit2014,Mich2025}. 

This consistency is significant considering Fig. \ref{fig:Mass}, showing a {strong positive correlation} between $E_k$ and $M_{^{56}Ni}$ for CC-SNe, extending from typical type II events such as SN 1987A to energetic GRB-SNe SN 1998bw, out to the {relativistic type Ic event SN 1999al with $M_{^{56}{\rm Ni}}\simeq 4M_\odot$ at $E_k\simeq 10^{53}$erg \citep{Iwamoto1998,Mario2003,Nomoto2006,Smartt2009b,Woosley2006,Umeda2008}}.
Notably, SN 2023ixf shares common properties with SN 1987A in energetics, non-relativistic ejecta, $^{56}$Ni production and, {likely}, a preferred axis of symmetry. 
\textcolor{black}{The location of SN 2023ixf close to SN 1987A in the $E_k$–$^{56}$Ni plane supports an interpretation in terms of a relatively standard Type II explosion, naturally favoring a NS central engine scenario.}{We note, however, that estimates of the synthesized 56Ni mass are model dependent and may vary among different analyses. Nevertheless, these measurements fall within the uncertainties of SN 1987A \citep{Seit2014}, maintaining the similarity between these two events. Accordingly, Fig. 3 should be interpreted primarily as an informative comparison rather than as a quantitative diagnostic.”}


{However, unlike SN 1987A, the inconclusive detection of a preferred axis in SN 2023ixf at these early stages does not decisively constrain angular momentum versus MeV-neutrino powered explosion mechanisms. While the initial explosion may be powered by a NS (spin-down and/or MeV-neutrino emission), the perhaps unlikely event of delayed gravitational collapse of a HMNS to a BH implies a possible but not necessarily secondary engine activity (BH spin-down and/or accretion). (Delayed gravitational collapse process is evidenced in GW170817B/GRB170817A.) {The non-detection of GW emission, combined with the observed ejecta energetics and moderate 56Ni production, appears more consistent with a NS-powered explosion than with an energetic BH-powered scenario.}}

Accordingly, these multi-messenger constraints comprising null results from direct gravitational-wave observations subject to {equation \eqref{EQN_threshold}} and photometric-spectroscopic data, present a remarkably robust pointer to the central engine of SN 2023ixf, most likely a proto-NS rather than a BH. 

\begin{table*}[]
\begin{tabular}{|c|c||c|c|c|}
\cline{1-5}
Mass
&Horizon Distance
& SN Ibc  
& LGRB (true)  
& LLGRB (true) \\ \hline \hline

$M_\odot$
&(Mpc)
& (yr)$^{-1}$ 
& (yr)$^{-1}$ 
&(yr)$^{-1}$ \\ \hline \hline

$M\simeq 5M_\odot$ & 100--200 Mpc
& $\sim 100$--$1000$ 
& $\sim 1$--$10$ 
& $\sim 10$--$100$ \\ \hline

$M\simeq 2.8M_\odot$ & 50--100 Mpc
& $\sim 10$--$100$ 
& $\sim 0.1$--few 
& $\sim 1$--$10$ \\ \hline

\end{tabular}
\caption{Central engines of CC-SNe are expected to be a NS or a BH with an inherent diversity in initial spin, the latter further including a diversity in mass. Shown are statistical estimates of observational opportunities to search for potential GW detection from spinning BH central engines of relatively more energetic CC-SNe Ibc and their offspring in GRBs. Here we differentiate between BH masses $5 M_\odot$ and $2.8M_\odot$, considering the associated horizon distance to either one. The case of $2.8M_\odot$ gives a conservative prospect for GW observations from CC-SNe. ‌The lower/higher horizon distance indicated refers to BHs with moderate/rapid initial spin, $\hat{a}\simeq 0.5$ versus $\hat{a}\simeq 0.8$ (based on Fig. \ref{fig:spectrum}). Consequently, the statistical event rates show vary by an order of magnitude. {Together with CC-SNe II, these event rates} will improve by a factor of a few \citep{mvp2024}. }
\label{T1}
\end{table*}

\section{Conclusion}
\label{S_conclusion}

We focused on a multi-messenger analysis of the recent {Type II CC-SNe SN~2023ixf}.
Given its proximity $D\simeq 6.85$\, Mpc, we include novel constraints from GW-observations using model-independet BMF analysis, during ER15 just before LIGO O4, to constrain its central engine and progenitor mass, complementing photometric-spectroscopic constraints.

Given the proximity of SN 2023ixf, a long duration spin-down GW signal of the type expected from a rotating black hole similar to GW170817B, would have been detectable in the available data (Fig. \ref{fig:BMF}). Our results show a non-detection of such signal. As for accretion powered gravitational radiation, this is disfavored based on constraints by $^{56}$Ni production associated with SN 2023ixf (\S \ref{S_discussion}). Within this framework, this non-detection (\S \ref{S_prerequisite}) disfavors BH central engine. Further considering the photometric-spectroscopic similarity between SN 2023ixf and SN 1987A, 
together with the preferred axis of SN 2023ixf 
\citep{Sing2024}, {our results show no evidence of BH formation in SN 2023ixf.
Instead, within the framework of multi-messenger astronomy, namely, GW170817B-like BH spin-down signals and $^{56}$Ni production, our results are more naturally consistent with the formation of an NS.} 


{This inference is remarkably consistent with independent EM estimates of the progenitor mass of SN 2023ixf, which typically fall in the range $\sim 10–15 M_\odot$ 
from hydrodynamical modeling and pre-explosion imaging (e.g. \cite{Sora2023, Bers2024,Ferr2024,VanDyk2024,Mori2024, Sing2024,Fang2024,Jaco2025,Jaco2025a,Hsu2025,Li2025,Ford2025,Kumar2025,Mich2025,Qin2024}, see also \S \ref{S_ixf}), none of which exceeds the first compactness peak \citep{Laplace2025}.}

The initial spin frequency of this newly born NS remnant may be below 50Hz, putting it in the category of young normal pulsars in the $P\text{—}\dot{P}$ diagram \citep{Taylor1977}.

Although uncertain, current event rates of CC-SNe, specially CC-CNe Ibc as the parent population of LGRBs and LLGRBs, give a prospect for observation of GWs from rotating BH central engines within the estimated horizon distance (\S \ref{S_discussion}). Table \ref{T1} shows this prospect for the typical BH masses of $5 M_\odot$ and, alternatively, $2.8 M_\odot$. While the central engine of CC-SNe are more likely to be $>5 M_\odot$, the case of $2.8 M_\odot$ gives a conservative view on the observing time to detect spin-down GW signals.

\section*{Acknowledgment}
{We thank the anonymous reviewer for constructive comments.} The authors gratefully acknowledge support from the National Research Foundation of Korea under grant number
RS-2024-00334550. This work made use of LIGO ER15 data from the LIGO Open Science Center provided by the LIGO Laboratory and LIGO Scientific Collaboration. LIGO is funded by the U.S. National Science Foundation.

\bibliography{IXF-bib}{}

@ARTICLE{van11,
       author = {{van~Putten}, M.~H.~P.~M. and {Della Valle}, M. and {Levinson}, A.},
        title = "{Electromagnetic priors for black hole spindown in searches for gravitational waves from supernovae and long GRBs}",
      journal = {\aap},
         year = 2011,
        month = nov,
       volume = {535},
          eid = {L6},
        pages = {L6},
          doi = {10.1051/0004-6361/201118080},
archivePrefix = {arXiv},
       eprint = {1111.0137},
 primaryClass = {astro-ph.HE},
       adsurl = {https://ui.adsabs.harvard.edu/abs/2011A&A...535L...6V}
}

@ARTICLE{cos25,
       author = {{Cosentino}, Stefano P. and {Pumo}, Maria L. and {Cherubini}, Silvio},
        title = "{High-energy neutrinos by hydrogen-rich supernovae interacting with low-massive circumstellar medium: the case of SN 2023ixf}",
      journal = {\mnras},
         year = 2025,
        month = jul,
       volume = {540},
       number = {4},
        pages = {2894-2913},
          doi = {10.1093/mnras/staf861},
archivePrefix = {arXiv},
       eprint = {2503.03699},
 primaryClass = {astro-ph.HE},
       adsurl = {https://ui.adsabs.harvard.edu/abs/2025MNRAS.540.2894C}
}

@ARTICLE{fen12,
       author = {{Fender}, Rob and {Belloni}, Tomaso},
        title = "{Stellar-Mass Black Holes and Ultraluminous X-ray Sources}",
      journal = {Science},
         year = 2012,
        month = aug,
       volume = {337},
       number = {6094},
        pages = {540},
          doi = {10.1126/science.1221790},
archivePrefix = {arXiv},
       eprint = {1208.1138},
 primaryClass = {astro-ph.HE},
       adsurl = {https://ui.adsabs.harvard.edu/abs/2012Sci...337..540F}
}

@article{van25D,
  author    = {van~Putten, Maurice H. P. M. and Wilson, Leighton and Lavely, Adam and Hair, Mark},
  title     = {Slide {FFT} on a homogeneous mesh in wafer-scale computing},
  journal   = {Discover Computing},
  volume    = {28},
  pages     = {16},
  year      = {2025},
  doi       = {10.1007/s10791-025-09508-2},
  url       = {https://link.springer.com/article/10.1007/s10791-025-09508-2}
}

@ARTICLE{van16A,
       author = {{van~Putten}, Maurice H.~P.~M.},
        title = "{Directed Searches for Broadband Extended Gravitational Wave Emission in Nearby Energetic Core-collapse Supernovae}",
      journal = {\apj},
         year = 2016,
        month = mar,
       volume = {819},
       number = {2},
          eid = {169},
        pages = {169},
          doi = {10.3847/0004-637X/819/2/169},
archivePrefix = {arXiv},
       eprint = {1602.03634},
 primaryClass = {astro-ph.HE},
       adsurl = {https://ui.adsabs.harvard.edu/abs/2016ApJ...819..169V}
}

@article{fal04,
  author  = {Falcke, Heino and K{\"o}rding, Elmar and Markoff, Sera},
  title   = {A scheme to unify low-power accreting black holes},
  journal = {Astronomy \& Astrophysics},
  volume  = {414},
  pages   = {895--903},
  year    = {2004}
}

@ARTICLE{fen03,
       author = {{Fender}, R.~P. and {Gallo}, E. and {Jonker}, P.~G.},
        title = "{Jet-dominated states: an alternative to advection across black hole event horizons in `quiescent' X-ray binaries}",
      journal = {\mnras},
         year = 2003,
        month = aug,
       volume = {343},
       number = {4},
        pages = {L99-L103},
          doi = {10.1046/j.1365-8711.2003.06950.x},
archivePrefix = {arXiv},
       eprint = {astro-ph/0306614},
 primaryClass = {astro-ph},
       adsurl = {https://ui.adsabs.harvard.edu/abs/2003MNRAS.343L..99F}
}

@ARTICLE{mer03,
       author = {{Merloni}, Andrea and {Heinz}, Sebastian and {di Matteo}, Tiziana},
        title = "{A Fundamental Plane of black hole activity}",
      journal = {\mnras},
         year = 2003,
        month = nov,
       volume = {345},
       number = {4},
        pages = {1057-1076},
          doi = {10.1046/j.1365-2966.2003.07017.x},
archivePrefix = {arXiv},
       eprint = {astro-ph/0305261},
 primaryClass = {astro-ph},
       adsurl = {https://ui.adsabs.harvard.edu/abs/2003MNRAS.345.1057M}
}

@article{mvp2024,
  title = {Unveiling the Central Engine of Core-collapse Supernovae in the Local Universe: Neutron Star or Black Hole?},
  volume = {972},
  ISSN = {2041-8213},
  url = {http://dx.doi.org/10.3847/2041-8213/ad710f},
  DOI = {10.3847/2041-8213/ad710f},
  number = {2},
  journal = {The Astrophysical Journal Letters},
  publisher = {American Astronomical Society},
  author = {van~Putten,  Maurice H. P. M. and Aghaei Abchouyeh,  Maryam and Della Valle,  Massimo},
  year = {2024},
  month = sep,
  pages = {L23}
}

@ARTICLE{Graham2019,
       author = {{Graham}, Matthew J. and {Kulkarni}, S.~R. and {Bellm}, Eric C. and {Adams}, Scott M. and {Barbarino}, Cristina and {Blagorodnova}, Nadejda and {Bodewits}, Dennis and {Bolin}, Bryce and {Brady}, Patrick R. and {Cenko}, S. Bradley and {Chang}, Chan-Kao and {Coughlin}, Michael W. and {De}, Kishalay and {Eadie}, Gwendolyn and {Farnham}, Tony L. and {Feindt}, Ulrich and {Franckowiak}, Anna and {Fremling}, Christoffer and {Gezari}, Suvi and {Ghosh}, Shaon and {Goldstein}, Daniel A. and {Golkhou}, V. Zach and {Goobar}, Ariel and {Ho}, Anna Y.~Q. and {Huppenkothen}, Daniela and {Ivezi{\'c}}, {\v{Z}}eljko and {Jones}, R. Lynne and {Juric}, Mario and {Kaplan}, David L. and {Kasliwal}, Mansi M. and {Kelley}, Michael S.~P. and {Kupfer}, Thomas and {Lee}, Chien-De and {Lin}, Hsing Wen and {Lunnan}, Ragnhild and {Mahabal}, Ashish A. and {Miller}, Adam A. and {Ngeow}, Chow-Choong and {Nugent}, Peter and {Ofek}, Eran O. and {Prince}, Thomas A. and {Rauch}, Ludwig and {van Roestel}, Jan and {Schulze}, Steve and {Singer}, Leo P. and {Sollerman}, Jesper and {Taddia}, Francesco and {Yan}, Lin and {Ye}, Quan-Zhi and {Yu}, Po-Chieh and {Barlow}, Tom and {Bauer}, James and {Beck}, Ron and {Belicki}, Justin and {Biswas}, Rahul and {Brinnel}, Valery and {Brooke}, Tim and {Bue}, Brian and {Bulla}, Mattia and {Burruss}, Rick and {Connolly}, Andrew and {Cromer}, John and {Cunningham}, Virginia and {Dekany}, Richard and {Delacroix}, Alex and {Desai}, Vandana and {Duev}, Dmitry A. and {Feeney}, Michael and {Flynn}, David and {Frederick}, Sara and {Gal-Yam}, Avishay and {Giomi}, Matteo and {Groom}, Steven and {Hacopians}, Eugean and {Hale}, David and {Helou}, George and {Henning}, John and {Hover}, David and {Hillenbrand}, Lynne A. and {Howell}, Justin and {Hung}, Tiara and {Imel}, David and {Ip}, Wing-Huen and {Jackson}, Edward and {Kaspi}, Shai and {Kaye}, Stephen and {Kowalski}, Marek and {Kramer}, Emily and {Kuhn}, Michael and {Landry}, Walter and {Laher}, Russ R. and {Mao}, Peter and {Masci}, Frank J. and {Monkewitz}, Serge and {Murphy}, Patrick and {Nordin}, Jakob and {Patterson}, Maria T. and {Penprase}, Bryan and {Porter}, Michael and {Rebbapragada}, Umaa and {Reiley}, Dan and {Riddle}, Reed and {Rigault}, Mickael and {Rodriguez}, Hector and {Rusholme}, Ben and {van Santen}, Jakob and {Shupe}, David L. and {Smith}, Roger M. and {Soumagnac}, Maayane T. and {Stein}, Robert and {Surace}, Jason and {Szkody}, Paula and {Terek}, Scott and {Van Sistine}, Angela and {van Velzen}, Sjoert and {Vestrand}, W. Thomas and {Walters}, Richard and {Ward}, Charlotte and {Zhang}, Chaoran and {Zolkower}, Jeffry},
        title = "{The Zwicky Transient Facility: Science Objectives}",
      journal = {\pasp},
         year = 2019,
        month = jul,
       volume = {131},
       number = {1001},
        pages = {078001},
          doi = {10.1088/1538-3873/ab006c},
archivePrefix = {arXiv},
       eprint = {1902.01945},
 primaryClass = {astro-ph.IM},
       adsurl = {https://ui.adsabs.harvard.edu/abs/2019PASP..131g8001G}
}

@article{Aleo2023,
  title = {The Young Supernova Experiment Data Release 1 (YSE DR1): Light Curves and Photometric Classification of 1975 Supernovae},
  volume = {266},
  ISSN = {1538-4365},
  url = {http://dx.doi.org/10.3847/1538-4365/acbfba},
  DOI = {10.3847/1538-4365/acbfba},
  number = {1},
  journal = {The Astrophysical Journal Supplement Series},
  publisher = {American Astronomical Society},
  author = {Aleo,  P. D. and Malanchev,  K. and Sharief,  S. and Jones,  D. O. and Narayan,  G. and Foley,  R. J. and Villar,  V. A. and Angus,  C. R. and Baldassare,  V. F. and Bustamante-Rosell,  M. J. and Chatterjee,  D. and Cold,  C. and Coulter,  D. A. and Davis,  K. W. and Dhawan,  S. and Drout,  M. R. and Engel,  A. and French,  K. D. and Gagliano,  A. and Gall,  C. and Hjorth,  J. and Huber,  M. E. and Jacobson-Galán,  W. V. and Kilpatrick,  C. D. and Langeroodi,  D. and Macias,  P. and Mandel,  K. S. and Margutti,  R. and Matasić,  F. and McGill,  P. and Pierel,  J. D. R. and Ramirez-Ruiz,  E. and Ransome,  C. L. and Rojas-Bravo,  C. and Siebert,  M. R. and Smith,  K. W. and de Soto,  K. M. and Stroh,  M. C. and Tinyanont,  S. and Taggart,  K. and Ward,  S. M. and Wojtak,  R. and Auchettl,  K. and Blanchard,  P. K. and de Boer,  T. J. L. and Boyd,  B. M. and Carroll,  C. M. and Chambers,  K. C. and DeMarchi,  L. and Dimitriadis,  G. and Dodd,  S. A. and Earl,  N. and Farias,  D. and Gao,  H. and Gomez,  S. and Grayling,  M. and Grillo,  C. and Hayes,  E. E. and Hung,  T. and Izzo,  L. and Khetan,  N. and Kolborg,  A. N. and Law-Smith,  J. A. P. and LeBaron,  N. and Lin,  C.-C. and Luo,  Y. and Magnier,  E. A. and Matthews,  D. and Mockler,  B. and O’Grady,  A. J. G. and Pan,  Y.-C. and Politsch,  C. A. and Raimundo,  S. I. and Rest,  A. and Ridden-Harper,  R. and Sarangi,  A. and Schrøder,  S. L. and Smartt,  S. J. and Terreran,  G. and Thorp,  S. and Vazquez,  J. and Wainscoat,  R. J. and Wang,  Q. and Wasserman,  A. R. and Yadavalli,  S. K. and Yarza,  R. and Zenati,  Y.},
  year = {2023},
  month = may,
  pages = {9}
}

@article{Tonry2018,
  title = {ATLAS: A High-cadence All-sky Survey System},
  volume = {130},
  ISSN = {1538-3873},
  url = {http://dx.doi.org/10.1088/1538-3873/aabadf},
  DOI = {10.1088/1538-3873/aabadf},
  number = {988},
  journal = {Publications of the Astronomical Society of the Pacific},
  publisher = {IOP Publishing},
  author = {Tonry,  J. L. and Denneau,  L. and Heinze,  A. N. and Stalder,  B. and Smith,  K. W. and Smartt,  S. J. and Stubbs,  C. W. and Weiland,  H. J. and Rest,  A.},
  year = {2018},
  month = may,
  pages = {064505}
}

@ARTICLE{Lipunov2007,
       author = {{Lipunov}, V.~M. and {Kornilov}, V.~G. and {Krylov}, A.~V. and {Tyurina}, N.~V. and {Belinskii}, A.~A. and {Gorbovskoi}, E.~S. and {Kuvshinov}, D.~A. and {Gritsyk}, P.~A. and {Antipov}, G.~A. and {Borisov}, G.~V. and {Sankovich}, A.~V. and {Vladimirov}, V.~V. and {Vybornov}, V.~I. and {Kuznetsov}, A.~S.},
        title = "{Optical observations of gamma-ray bursts, the discovery of supernovae 2005bv, 2005ee, and 2006ak, and searches for transients using the ``MASTER'' robotic telescope}",
      journal = {Astronomy Reports},
         year = 2007,
        month = dec,
       volume = {51},
       number = {12},
        pages = {1004-1025},
          doi = {10.1134/S1063772907120050},
archivePrefix = {arXiv},
       eprint = {0711.0037},
 primaryClass = {astro-ph},
       adsurl = {https://ui.adsabs.harvard.edu/abs/2007ARep...51.1004L}
}

@article{Jones2021,
  title = {The Young Supernova Experiment: Survey Goals,  Overview,  and Operations},
  volume = {908},
  ISSN = {1538-4357},
  url = {http://dx.doi.org/10.3847/1538-4357/abd7f5},
  DOI = {10.3847/1538-4357/abd7f5},
  number = {2},
  journal = {The Astrophysical Journal},
  publisher = {American Astronomical Society},
  author = {Jones,  D. O. and Foley,  R. J. and Narayan,  G. and Hjorth,  J. and Huber,  M. E. and Aleo,  P. D. and Alexander,  K. D. and Angus,  C. R. and Auchettl,  K. and Baldassare,  V. F. and Bruun,  S. H. and Chambers,  K. C. and Chatterjee,  D. and Coppejans,  D. L. and Coulter,  D. A. and DeMarchi,  L. and Dimitriadis,  G. and Drout,  M. R. and Engel,  A. and French,  K. D. and Gagliano,  A. and Gall,  C. and Hung,  T. and Izzo,  L. and Jacobson-Galán,  W. V. and Kilpatrick,  C. D. and Korhonen,  H. and Margutti,  R. and Raimundo,  S. I. and Ramirez-Ruiz,  E. and Rest,  A. and Rojas-Bravo,  C. and Siebert,  M. R. and Smartt,  S. J. and Smith,  K. W. and Terreran,  G. and Wang,  Q. and Wojtak,  R. and Agnello,  A. and Ansari,  Z. and Arendse,  N. and Baldeschi,  A. and Blanchard,  P. K. and Brethauer,  D. and Bright,  J. S. and Brown,  J. S. and Boer,  T. J. L. de and Dodd,  S. A. and Fairlamb,  J. R. and Grillo,  C. and Hajela,  A. and Cold,  C. and Kolborg,  A. N. and Law-Smith,  J. A. P. and Lin,  C.-C. and Magnier,  E. A. and Malanchev,  K. and Matthews,  D. and Mockler,  B. and Muthukrishna,  D. and Pan,  Y.-C. and Pfister,  H. and Ramanah,  D. K. and Rest,  S. and Sarangi,  A. and Schrøder,  S. L. and Stauffer,  C. and Stroh,  M. C. and Taggart,  K. L. and Tinyanont,  S. and Wainscoat,  R. J.},
  year = {2021},
  month = feb,
  pages = {143}
}

@article{Pejcha2015,
  title = {THE LANDSCAPE OF THE NEUTRINO MECHANISM OF CORE-COLLAPSE SUPERNOVAE: NEUTRON STAR AND BLACK HOLE MASS FUNCTIONS,  EXPLOSION ENERGIES,  AND NICKEL YIELDS},
  volume = {801},
  ISSN = {1538-4357},
  url = {http://dx.doi.org/10.1088/0004-637X/801/2/90},
  DOI = {10.1088/0004-637x/801/2/90},
  number = {2},
  journal = {The Astrophysical Journal},
  publisher = {American Astronomical Society},
  author = {Pejcha,  Ondřej and Thompson,  Todd A.},
  year = {2015},
  month = mar,
  pages = {90}
}

@article{Oberg2020,
  title = {Magnetorotational core collapse of possible GRB progenitors – I. Explosion mechanisms},
  volume = {492},
  ISSN = {1365-2966},
  url = {http://dx.doi.org/10.1093/mnras/staa096},
  DOI = {10.1093/mnras/staa096},
  number = {4},
  journal = {Monthly Notices of the Royal Astronomical Society},
  publisher = {Oxford University Press (OUP)},
  author = {Obergaulinger,  M and Aloy,  M {\'A} },
  year = {2020},
  month = jan,
  pages = {4613–4634}
}

@article{Sawai2014,
  title = {INFLUENCE OF MAGNETOROTATIONAL INSTABILITY ON NEUTRINO HEATING: A NEW MECHANISM FOR WEAKLY MAGNETIZED CORE-COLLAPSE SUPERNOVAE},
  volume = {784},
  ISSN = {2041-8213},
  url = {http://dx.doi.org/10.1088/2041-8205/784/1/L10},
  DOI = {10.1088/2041-8205/784/1/l10},
  number = {1},
  journal = {The Astrophysical Journal},
  publisher = {American Astronomical Society},
  author = {Sawai,  Hidetomo and Yamada,  Shoichi},
  year = {2014},
  month = mar,
  pages = {L10}
}

@article{Janka2012,
  title = {Explosion Mechanisms of Core-Collapse Supernovae},
  volume = {62},
  ISSN = {1545-4134},
  url = {http://dx.doi.org/10.1146/annurev-nucl-102711-094901},
  DOI = {10.1146/annurev-nucl-102711-094901},
  number = {1},
  journal = {Annual Review of Nuclear and Particle Science},
  publisher = {Annual Reviews},
  author = {Janka,  Hans-Thomas},
  year = {2012},
  month = nov,
  pages = {407–451}
}

@article{Papish2015,
  title = {A call for a paradigm shift from neutrino-driven to jet-driven core-collapse supernova mechanisms},
  volume = {448},
  ISSN = {0035-8711},
  url = {http://dx.doi.org/10.1093/mnras/stv131},
  DOI = {10.1093/mnras/stv131},
  number = {3},
  journal = {Monthly Notices of the Royal Astronomical Society},
  publisher = {Oxford University Press (OUP)},
  author = {Papish,  Oded and Nordhaus,  Jason and Soker,  Noam},
  year = {2015},
  month = feb,
  pages = {2362–2367}
}

@article{Smartt2009,
  title = {The death of massive stars - I. Observational constraints on the progenitors of Type II-P supernovae},
  volume = {395},
  ISSN = {1365-2966},
  url = {http://dx.doi.org/10.1111/j.1365-2966.2009.14506.x},
  DOI = {10.1111/j.1365-2966.2009.14506.x},
  number = {3},
  journal = {Monthly Notices of the Royal Astronomical Society},
  publisher = {Oxford University Press (OUP)},
  author = {Smartt,  S. J. and Eldridge,  J. J. and Crockett,  R. M. and Maund,  J. R.},
  year = {2009},
  month = may,
  pages = {1409–1437}
}

@article{Sukhb2016,
  title = {CORE-COLLAPSE SUPERNOVAE FROM 9 TO 120 SOLAR MASSES BASED ON NEUTRINO-POWERED EXPLOSIONS},
  volume = {821},
  ISSN = {1538-4357},
  url = {http://dx.doi.org/10.3847/0004-637X/821/1/38},
  DOI = {10.3847/0004-637x/821/1/38},
  number = {1},
  journal = {The Astrophysical Journal},
  publisher = {American Astronomical Society},
  author = {Sukhbold,  Tuguldur and Ertl,  T. and Woosley,  S. E. and Brown,  Justin M. and Janka,  H.-T.},
  year = {2016},
  month = apr,
  pages = {38}
}

@article{Iwamoto1998,
  title = {A hypernova model for the supernova associated with the γ-ray burst of 25 April 1998},
  volume = {395},
  ISSN = {1476-4687},
  url = {http://dx.doi.org/10.1038/27155},
  DOI = {10.1038/27155},
  number = {6703},
  journal = {Nature},
  publisher = {Springer Science and Business Media LLC},
  author = {Iwamoto,  K. and Mazzali,  P. A. and Nomoto,  K. and Umeda,  H. and Nakamura,  T. and Patat,  F. and Danziger,  I. J. and Young,  T. R. and Suzuki,  T. and Shigeyama,  T. and Augusteijn,  T. and Doublier,  V. and Gonzalez,  J.-F. and Boehnhardt,  H. and Brewer,  J. and Hainaut,  O. R. and Lidman,  C. and Leibundgut,  B. and Cappellaro,  E. and Turatto,  M. and Galama,  T. J. and Vreeswijk,  P. M. and Kouveliotou,  C. and van Paradijs,  J. and Pian,  E. and Palazzi,  E. and Frontera,  F.},
  year = {1998},
  month = oct,
  pages = {672–674}
}

@article{Woosley2006,
  title = {The Supernova–Gamma-Ray Burst Connection},
  volume = {44},
  ISSN = {1545-4282},
  url = {http://dx.doi.org/10.1146/annurev.astro.43.072103.150558},
  DOI = {10.1146/annurev.astro.43.072103.150558},
  number = {1},
  journal = {Annual Review of Astronomy and Astrophysics},
  publisher = {Annual Reviews},
  author = {Woosley,  S.E. and Bloom,  J.S.},
  year = {2006},
  month = sep,
  pages = {507–556}
}

@ARTICLE{Alp2018,
       author = {{Alp}, Dennis and {Larsson}, Josefin and {Fransson}, Claes and {Indebetouw}, Remy and {Jerkstrand}, Anders and {Ahola}, Antero and {Burrows}, David and {Challis}, Peter and {Cigan}, Phil and {Cikota}, Aleksandar and {Kirshner}, Robert P. and {van Loon}, Jacco Th. and {Mattila}, Seppo and {Ng}, C.-Y. and {Park}, Sangwook and {Spyromilio}, Jason and {Woosley}, Stan and {Baes}, Maarten and {Bouchet}, Patrice and {Chevalier}, Roger and {Frank}, Kari A. and {Gaensler}, B.~M. and {Gomez}, Haley and {Janka}, Hans-Thomas and {Leibundgut}, Bruno and {Lundqvist}, Peter and {Marcaide}, Jon and {Matsuura}, Mikako and {Sollerman}, Jesper and {Sonneborn}, George and {Staveley-Smith}, Lister and {Zanardo}, Giovanna and {Gabler}, Michael and {Taddia}, Francesco and {Wheeler}, J. Craig},
        title = "{The 30 Year Search for the Compact Object in SN 1987A}",
      journal = {ApJ},
         year = 2018,
        month = sep,
       volume = {864},
       number = {2},
          eid = {174},
        pages = {174},
          doi = {10.3847/1538-4357/aad739},
archivePrefix = {arXiv},
       eprint = {1805.04526},
 primaryClass = {astro-ph.HE},
       adsurl = {https://ui.adsabs.harvard.edu/abs/2018ApJ...864..174A}
}

@ARTICLE{Fran2024,
       author = {{Fransson}, C. and {Barlow}, M.~J. and {Kavanagh}, P.~J. and {Larsson}, J. and {Jones}, O.~C. and {Sargent}, B. and {Meixner}, M. and {Bouchet}, P. and {Temim}, T. and {Wright}, G.~S. and {Blommaert}, J.~A.~D.~L. and {Habel}, N. and {Hirschauer}, A.~S. and {Hjorth}, J. and {Lenki{\'c}}, L. and {Tikkanen}, T. and {Wesson}, R. and {Coulais}, A. and {Fox}, O.~D. and {Gastaud}, R. and {Glasse}, A. and {Jaspers}, J. and {Krause}, O. and {Lau}, R.~M. and {Nayak}, O. and {Rest}, A. and {Colina}, L. and {van Dishoeck}, E.~F. and {G{\"u}del}, M. and {Henning}, Th. and {Lagage}, P.-O. and {{\"O}stlin}, G. and {Ray}, T.~P. and {Vandenbussche}, B.},
        title = "{Emission lines due to ionizing radiation from a compact object in the remnant of Supernova 1987A}",
      journal = {Science},
         year = 2024,
        month = feb,
       volume = {383},
       number = {6685},
        pages = {898-903},
          doi = {10.1126/science.adj5796},
archivePrefix = {arXiv},
       eprint = {2403.04386},
 primaryClass = {astro-ph.HE},
       adsurl = {https://ui.adsabs.harvard.edu/abs/2024Sci...383..898F}
}

@ARTICLE{mvp2023,
       author = {{van~Putten}, Maurice H.~P.~M. and {Della Valle}, Massimo},
        title = "{Central engine of GRB170817A: Neutron star versus Kerr black hole based on multimessenger calorimetry and event timing}",
      journal = {Astronomy \& Astrophysics},
         year = 2023,
        month = jan,
       volume = {669},
          eid = {A36},
        pages = {A36},
          doi = {10.1051/0004-6361/202142974},
archivePrefix = {arXiv},
       eprint = {2212.03295},
 primaryClass = {astro-ph.HE},
       adsurl = {https://ui.adsabs.harvard.edu/abs/2023A&A...669A..36V}
}

@ARTICLE{mvp2019,
       author = {van~Putten, Maurice H.~P.~M. and Della Valle, Massimo and Levinson, Amir},
        title = "{Multi-messenger Extended Emission from the Compact Remnant in GW170817}",
      journal = {The Astrophysical Journal Letters},
         year = 2019,
        month = may,
       volume = {876},
       number = {1},
          eid = {L2},
        pages = {L2},
          doi = {10.3847/2041-8213/ab18a2},
archivePrefix = {arXiv},
       eprint = {1910.12730},
 primaryClass = {astro-ph.HE},
       adsurl = {https://ui.adsabs.harvard.edu/abs/2019ApJ...876L...2V}
}

@ARTICLE{Bis1970,
       author = {{Bisnovatyi-Kogan}, G.~S.},
        title = "{The Explosion of a Rotating Star As a Supernova Mechanism.}",
      journal = {\azh},
         year = 1970,
        month = aug,
       volume = {47},
        pages = {813},
       adsurl = {https://ui.adsabs.harvard.edu/abs/1970AZh....47..813B}
}

@ARTICLE{Bis1971,
       author = {{Bisnovatyi-Kogan}, G.~S.},
        title = "{The Explosion of a Rotating Star As a Supernova Mechanism.}",
      journal = {\sovast},
         year = 1971,
        month = feb,
       volume = {14},
        pages = {652},
       adsurl = {https://ui.adsabs.harvard.edu/abs/1971SvA....14..652B}
}

@ARTICLE{ligo2017,
       author = {{Abbott}, B.~P. and {Abbott}, R. and {Abbott}, T.~D. and {Acernese}, F. and {Ackley}, K. and {Adams}, C. and {Adams}, T. and {Addesso}, P. and {Adhikari}, R.~X. and {Adya}, V.~B. and {Affeldt}, C. and {Afrough}, M. and {Agarwal}, B. and {Agathos}, M. and {Agatsuma}, K. and {Aggarwal}, N. and {Aguiar}, O.~D. and {Aiello}, L. and {Ain}, A. and {Ajith}, P. and {Allen}, B. and {Allen}, G. and {Allocca}, A. and {Altin}, P.~A. and {Amato}, A. and {Ananyeva}, A. and {Anderson}, S.~B. and {Anderson}, W.~G. and {Angelova}, S.~V. and {Antier}, S. and {Appert}, S. and {Arai}, K. and {Araya}, M.~C. and {Areeda}, J.~S. and {Arnaud}, N. and {Arun}, K.~G. and {Ascenzi}, S. and {Ashton}, G. and {Ast}, M. and {Aston}, S.~M. and {Astone}, P. and {Atallah}, D.~V. and {Aufmuth}, P. and {Aulbert}, C. and {AultONeal}, K. and {Austin}, C. and {Avila-Alvarez}, A. and {Babak}, S. and {Bacon}, P. and {Bader}, M.~K.~M. and {Bae}, S. and {Baker}, P.~T. and {Baldaccini}, F. and {Ballardin}, G. and {Ballmer}, S.~W. and {Banagiri}, S. and {Barayoga}, J.~C. and {Barclay}, S.~E. and {Barish}, B.~C. and {Barker}, D. and {Barkett}, K. and {Barone}, F. and {Barr}, B. and {Barsotti}, L. and {Barsuglia}, M. and {Barta}, D. and {Barthelmy}, S.~D. and {Bartlett}, J. and {Bartos}, I. and {Bassiri}, R. and {Basti}, A. and {Batch}, J.~C. and {Bawaj}, M. and {Bayley}, J.~C. and {Bazzan}, M. and {B{\'e}csy}, B. and {Beer}, C. and {Bejger}, M. and {Belahcene}, I. and {Bell}, A.~S. and {Berger}, B.~K. and {Bergmann}, G. and {Bero}, J.~J. and {Berry}, C.~P.~L. and {Bersanetti}, D. and {Bertolini}, A. and {Betzwieser}, J. and {Bhagwat}, S. and {Bhandare}, R. and {Bilenko}, I.~A. and {Billingsley}, G. and {Billman}, C.~R. and {Birch}, J. and {Birney}, R. and {Birnholtz}, O. and {Biscans}, S. and {Biscoveanu}, S. and {Bisht}, A. and {Bitossi}, M. and {Biwer}, C. and {Bizouard}, M.~A. and {Blackburn}, J.~K. and {Blackman}, J. and {Blair}, C.~D. and {Blair}, D.~G. and {Blair}, R.~M. and {Bloemen}, S. and {Bock}, O. and {Bode}, N. and {Boer}, M. and {Bogaert}, G. and {Bohe}, A. and {Bondu}, F. and {Bonilla}, E. and {Bonnand}, R. and {Boom}, B.~A. and {Bork}, R. and {Boschi}, V. and {Bose}, S. and {Bossie}, K. and {Bouffanais}, Y. and {Bozzi}, A. and {Bradaschia}, C. and {Brady}, P.~R. and {Branchesi}, M. and {Brau}, J.~E. and {Briant}, T. and {Brillet}, A. and {Brinkmann}, M. and {Brisson}, V. and {Brockill}, P. and {Broida}, J.~E. and {Brooks}, A.~F. and {Brown}, D.~A. and {Brown}, D.~D. and {Brunett}, S. and {Buchanan}, C.~C. and {Buikema}, A. and {Bulik}, T. and {Bulten}, H.~J. and {Buonanno}, A. and {Buskulic}, D. and {Buy}, C. and {Byer}, R.~L. and {Cabero}, M. and {Cadonati}, L. and {Cagnoli}, G. and {Cahillane}, C. and {Calder{\'o}n Bustillo}, J. and {Callister}, T.~A. and {Calloni}, E. and {Camp}, J.~B. and {Canepa}, M. and {Canizares}, P. and {Cannon}, K.~C. and {Cao}, H. and {Cao}, J. and {Capano}, C.~D. and {Capocasa}, E. and {Carbognani}, F. and {Caride}, S. and {Carney}, M.~F. and {Casanueva Diaz}, J. and {Casentini}, C. and {Caudill}, S. and {Cavagli{\`a}}, M. and {Cavalier}, F. and {Cavalieri}, R. and {Cella}, G. and {Cepeda}, C.~B. and {Cerd{\'a}-Dur{\'a}n}, P. and {Cerretani}, G. and {Cesarini}, E. and {Chamberlin}, S.~J. and {Chan}, M. and {Chao}, S. and {Charlton}, P. and {Chase}, E. and {Chassande-Mottin}, E. and {Chatterjee}, D. and {Chatziioannou}, K. and {Cheeseboro}, B.~D. and {Chen}, H.~Y. and {Chen}, X. and {Chen}, Y. and {Cheng}, H.-P. and {Chia}, H. and {Chincarini}, A. and {Chiummo}, A. and {Chmiel}, T. and {Cho}, H.~S. and {Cho}, M. and {Chow}, J.~H. and {Christensen}, N. and {Chu}, Q. and {Chua}, A.~J.~K. and {Chua}, S. and {Chung}, A.~K.~W. and {Chung}, S. and {Ciani}, G.},
        title = "{Multi-messenger Observations of a Binary Neutron Star Merger}",
      journal = {\apjl},
         year = 2017,
        month = oct,
       volume = {848},
       number = {2},
          eid = {L12},
        pages = {L12},
          doi = {10.3847/2041-8213/aa91c9},
archivePrefix = {arXiv},
       eprint = {1710.05833},
 primaryClass = {astro-ph.HE},
       adsurl = {https://ui.adsabs.harvard.edu/abs/2017ApJ...848L..12A}
}

@ARTICLE{Hossein2023,
       author = {{Hosseinzadeh}, Griffin and {Farah}, Joseph and {Shrestha}, Manisha and {Sand}, David J. and {Dong}, Yize and {Brown}, Peter J. and {Bostroem}, K. Azalee and {Valenti}, Stefano and {Jha}, Saurabh W. and {Andrews}, Jennifer E. and {Arcavi}, Iair and {Haislip}, Joshua and {Hiramatsu}, Daichi and {Hoang}, Emily and {Howell}, D. Andrew and {Janzen}, Daryl and {Jencson}, Jacob E. and {Kouprianov}, Vladimir and {Lundquist}, Michael and {McCully}, Curtis and {Meza Retamal}, Nicolas E. and {Modjaz}, Maryam and {Newsome}, Megan and {Padilla Gonzalez}, Estefania and {Pearson}, Jeniveve and {Pellegrino}, Craig and {Ravi}, Aravind P. and {Reichart}, Daniel E. and {Smith}, Nathan and {Terreran}, Giacomo and {Vink{\'o}}, J{\'o}zsef},
        title = "{Shock Cooling and Possible Precursor Emission in the Early Light Curve of the Type II SN 2023ixf}",
      journal = {\apjl},
         year = 2023,
        month = aug,
       volume = {953},
       number = {1},
          eid = {L16},
        pages = {L16},
          doi = {10.3847/2041-8213/ace4c4},
archivePrefix = {arXiv},
       eprint = {2306.06097},
 primaryClass = {astro-ph.HE},
       adsurl = {https://ui.adsabs.harvard.edu/abs/2023ApJ...953L..16H}
}

@ARTICLE{Abbott2020,
       author = {{Abbott}, B.~P. and {Abbott}, R. and {Abbott}, T.~D. and {Abraham}, S. and {Acernese}, F. and {Ackley}, K. and {Adams}, C. and {Adya}, V.~B. and {Affeldt}, C. and {Agathos}, M. and {Agatsuma}, K. and {Aggarwal}, N. and {Aguiar}, O.~D. and {Aiello}, L. and {Ain}, A. and {Ajith}, P. and {Allen}, G. and {Allocca}, A. and {Aloy}, M.~A. and {Altin}, P.~A. and {Amato}, A. and {Anand}, S. and {Ananyeva}, A. and {Anderson}, S.~B. and {Anderson}, W.~G. and {Angelova}, S.~V. and {Antier}, S. and {Appert}, S. and {Arai}, K. and {Araya}, M.~C. and {Areeda}, J.~S. and {Ar{\`e}ne}, M. and {Arnaud}, N. and {Aronson}, S.~M. and {Ascenzi}, S. and {Ashton}, G. and {Aston}, S.~M. and {Astone}, P. and {Aubin}, F. and {Aufmuth}, P. and {AultONeal}, K. and {Austin}, C. and {Avendano}, V. and {Avila-Alvarez}, A. and {Babak}, S. and {Bacon}, P. and {Badaracco}, F. and {Bader}, M.~K.~M. and {Bae}, S. and {Baird}, J. and {Baker}, P.~T. and {Baldaccini}, F. and {Ballardin}, G. and {Ballmer}, S.~W. and {Bals}, A. and {Banagiri}, S. and {Barayoga}, J.~C. and {Barbieri}, C. and {Barclay}, S.~E. and {Barish}, B.~C. and {Barker}, D. and {Barkett}, K. and {Barnum}, S. and {Barone}, F. and {Barr}, B. and {Barsotti}, L. and {Barsuglia}, M. and {Barta}, D. and {Bartlett}, J. and {Bartos}, I. and {Bassiri}, R. and {Basti}, A. and {Bawaj}, M. and {Bayley}, J.~C. and {Bazzan}, M. and {B{\'e}csy}, B. and {Bejger}, M. and {Belahcene}, I. and {Bell}, A.~S. and {Beniwal}, D. and {Benjamin}, M.~G. and {Bergmann}, G. and {Bernuzzi}, S. and {Berry}, C.~P.~L. and {Bersanetti}, D. and {Bertolini}, A. and {Betzwieser}, J. and {Bhandare}, R. and {Bidler}, J. and {Biggs}, E. and {Bilenko}, I.~A. and {Bilgili}, S.~A. and {Billingsley}, G. and {Birney}, R. and {Birnholtz}, O. and {Biscans}, S. and {Bischi}, M. and {Biscoveanu}, S. and {Bisht}, A. and {Bitossi}, M. and {Bizouard}, M.~A. and {Blackburn}, J.~K. and {Blackman}, J. and {Blair}, C.~D. and {Blair}, D.~G. and {Blair}, R.~M. and {Bloemen}, S. and {Bobba}, F. and {Bode}, N. and {Boer}, M. and {Boetzel}, Y. and {Bogaert}, G. and {Bondu}, F. and {Bonnand}, R. and {Booker}, P. and {Boom}, B.~A. and {Bork}, R. and {Boschi}, V. and {Bose}, S. and {Bossilkov}, V. and {Bosveld}, J. and {Bouffanais}, Y. and {Bozzi}, A. and {Bradaschia}, C. and {Brady}, P.~R. and {Bramley}, A. and {Branchesi}, M. and {Brau}, J.~E. and {Breschi}, M. and {Briant}, T. and {Briggs}, J.~H. and {Brighenti}, F. and {Brillet}, A. and {Brinkmann}, M. and {Brockill}, P. and {Brooks}, A.~F. and {Brooks}, J. and {Brown}, D.~D. and {Brunett}, S. and {Buikema}, A. and {Bulik}, T. and {Bulten}, H.~J. and {Buonanno}, A. and {Buskulic}, D. and {Buy}, C. and {Byer}, R.~L. and {Cabero}, M. and {Cadonati}, L. and {Cagnoli}, G. and {Cahillane}, C. and {Bustillo}, J. Calder{\'o}n and {Callister}, T.~A. and {Calloni}, E. and {Camp}, J.~B. and {Campbell}, W.~A. and {Canepa}, M. and {Cannon}, K.~C. and {Cao}, H. and {Cao}, J. and {Carapella}, G. and {Carbognani}, F. and {Caride}, S. and {Carney}, M.~F. and {Carullo}, G. and {Diaz}, J. Casanueva and {Casentini}, C. and {Caudill}, S. and {Cavagli{\`a}}, M. and {Cavalier}, F. and {Cavalieri}, R. and {Cella}, G. and {Cerd{\'a}-Dur{\'a}n}, P. and {Cesarini}, E. and {Chaibi}, O. and {Chakravarti}, K. and {Chamberlin}, S.~J. and {Chan}, M. and {Chao}, S. and {Charlton}, P. and {Chase}, E.~A. and {Chassande-Mottin}, E. and {Chatterjee}, D. and {Chaturvedi}, M. and {Cheeseboro}, B.~D. and {Chen}, H.~Y. and {Chen}, X. and {Chen}, Y. and {Cheng}, H.-P. and {Cheong}, C.~K. and {Chia}, H.~Y. and {Chiadini}, F. and {Chincarini}, A. and {Chiummo}, A. and {Cho}, G. and {Cho}, H.~S. and {Cho}, M. and {Christensen}, N. and {Chu}, Q. and {Chua}, S. and {Chung}, K.~W.},
        title = "{Optically targeted search for gravitational waves emitted by core-collapse supernovae during the first and second observing runs of advanced LIGO and advanced Virgo}",
      journal = {\prd},
         year = 2020,
        month = apr,
       volume = {101},
       number = {8},
          eid = {084002},
        pages = {084002},
          doi = {10.1103/PhysRevD.101.084002},
archivePrefix = {arXiv},
       eprint = {1908.03584},
 primaryClass = {astro-ph.HE},
       adsurl = {https://ui.adsabs.harvard.edu/abs/2020PhRvD.101h4002A}
}
\bibliographystyle{aasjournal}

\end{document}